# Molecule-Induced Conformational Change in Boron Nitride Nanosheets with Enhanced Surface Adsorption

*Qiran Cai,[1] Aijun Du,[2] Guoping Gao,[2] Srikanth Mateti,[1] Bruce C.C. Cowie,[3] Dong Qian,[4] Shuang Zhang,[5] Yuerui Lu,[5] Lan Fu,[6] Takashi Taniguchi,[7] Shaoming Huang,[8] Ying Chen,[1]\* Rodney S. Ruoff[9,10,11] and Lu Hua Li[1]\**

1. Institute for Frontier Materials, Deakin University, 75 Pigdons Road, Waurn Ponds 3216, VIC (Australia)

2. School of Chemistry, Physics and Mechanical Engineering, Queensland University of Technology, 2 George St, Brisbane City, QLD 4000, Australia

3. Australian Synchrotron, 800 Blackburn Road, Clayton, VIC 3168, Australia

4. Department of Mechanical Engineering, The University of Texas at Dallas, Richardson, TX 75080, USA

5. Research School of Engineering, College of Engineering and Computer Science, The Australian National University, Canberra, ACT 2601, Australia

6. Department of Electronic Materials Engineering, Research School of Physics and Engineering, The Australian National University, Canberra, ACT 2601, Australia

7. National Institute for Materials Science, Namiki 1-1, Tsukuba, Ibaraki 305-0044, Japan

8. Nanomaterials and Chemistry Key Laboratory, Wenzhou University, 276 Xueyuan Middle Road, Wenzhou, Zhejiang 325027, China

9. IBS Center for Multidimensional Carbon Materials, Ulsan National Institute of Science and Technology, Ulsan 689-798, Republic of Korea.

10. Department of Chemistry, Ulsan National Institute of Science and Technology (UNIST), Ulsan 44919, Republic of Korea

11. School of Materials Science and Engineering, Ulsan National Institute of Science and Technology (UNIST), Ulsan 44919, Republic of Korea

E-mail: luhua.li@deakin.edu.au; ian.chen@deakin.edu.au

Surface interaction is extremely important to both fundamental research and practical application. Physisorption can induce shape and structural distortion (i.e. conformational changes) in macromolecular and biomolecular adsorbates, but such phenomenon has rarely been observed on adsorbents. Here, we demonstrate theoretically and experimentally that atomically thin boron nitride (BN) nanosheets as an adsorbent experience conformational changes upon surface adsorption of molecules, increasing adsorption energy and efficiency. The study not only provides new perspectives on the strong adsorption capability of BN nanosheets and many other two-dimensional nanomaterials but also opens up possibilities for many novel applications. For example, we demonstrate that BN nanosheets with the same surface area as bulk hBN particles are more effective in purification and sensing.

**1. Introduction**

Surface adsorption is a ubiquitous phenomenon vital to many physical, chemical and biological fields, and widely used in industry, such as purification, catalysis, sensor, chromatography, cell growth and drug delivery. Macromolecules and biomolecules, such as protein, ligand, peptide and DNA, have been found to have a special adsorption behavior, i.e. distortion of their shapes and structures after physisorbed on a surface.[1] Such phenomenon is called conformational change. Conformational change increases adsorption energy and greatly affects the chemical and biological activities of these molecules.[2] However, conformational change has been rarely observed on adsorbents. The relatively large rigidity of all traditional adsorbents, such as activated carbon, porous alumina, and zeolites, prevents them from interfacial interaction-induced deformation.

Boron nitride (BN) nanosheets, atomically thin layers of hexagonal BN (hBN), exhibit excellent surface adsorption of different molecules, and, therefore, are valuable for water cleaning, catalysis, sensing, etc.[3] Detailed studies revealing their unique adsorption behavior and mechanism are highly desirable. In contrast to traditional adsorbents, BN nanosheets have small bending moduli[4] and may experience conformational changes due to physisorption of molecules. Such behavior should improve their surface adsorption, and hence have implications for the development of novel applications.[5] However, there has been no theoretical or experimental study on molecule-induced conformational changes in BN nanosheets and their effect on their surface adsorption. Furthermore, it also lacks a straightforward experimental method to overcome the difficulty of detecting morphological changes at the atomic level after coverage of adsorbates.[6]

Raman spectroscopy is a well-accepted tool to study thickness,[7] crystallinity,[8] doping,[9] strain,[10] and lattice temperature[11] of graphene and BN nanosheets.[12] There has been a large number of experimental studies on the effect of doping on graphene and its Raman spectrum after molecule adsorption.[9d-f, 10h, 13] Unlike graphene, physisorbed molecules do not introduce doping to electrically insulating BN nanosheets or affect their Raman spectrum according to previous theoretical and experimental studies.[9d, 14] Nevertheless, Raman has never been used to detect conformational changes in 2D nanomaterials due to molecule adsorption.

Here, we report that atomically thin BN nanosheets bend or curve to better accommodate two model molecules, i.e. rhodamine 6G (R6G) and 4-mercaptobenzoic acid (4-MBA), and Raman spectroscopy is a straightforward and routine technique to detect such conformational changes. Conformational changes in BN nanosheets lead to stronger interfacial interaction and better surface adsorption capability than bulk crystals. Thus, for the same surface area, atomically thin BN nanosheets are more efficient in adsorbing molecules, and hence better candidates for water purification, sensing and many other applications related to surface adsorption. More importantly, such unique surface interaction and adsorption behavior should not be exclusive to BN nanosheets but general to many other two-dimensional (2D) nanomaterials.

## 2. Results and discussion

First, we carried out density functional theory (DFT) calculations to reveal the interaction between atomically thin BN nanosheets to R6G, a common analyte molecule. Figure 1a and b show side and top views of a R6G molecule on a flat and freestanding BN nanosheet. R6G preferred the so-called lying-down position with the xanthene ring structure parallel to the BN plane, with a distance about 0.8 nm. Such physical adsorption involved strong π-π interactions between R6G and BN. Intriguingly, this state was not stable. As shown in Figure 1d and e, the high flexibility of BN nanosheet made it spontaneously curved or distorted to alleviate the stress caused by the molecule adsorption. The deformation can be better seen from the height mapping in Figure 1f. The corresponding mappings of strain e11, e22, and e12 caused by R6G adsorption are shown in Figure 1g-i, and the average macro strain was $-0.16\%$ (compressive). The fixed BN nanosheet (Figure 1a-c) can mimic bulk hBN crystals, which are rigid and non-deformable upon adsorption. Conformational change in an adsorbate results in higher adsorption energy, as does conformational change in an adsorbent. Thus, atomically thin BN nanosheets should show higher adsorption energy and efficiency than bulk hBN which is not able to experience conformtional change, and such property is independent of surface area.

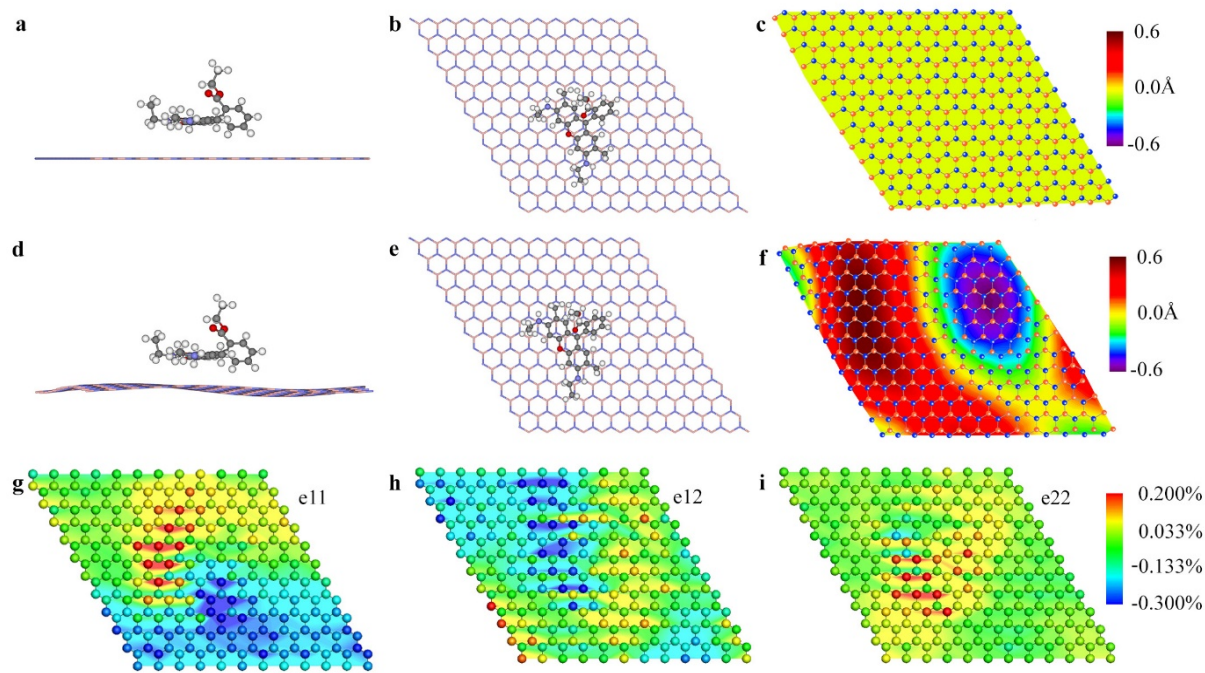

**Figure 1.** (a-c) Side and top views of a R6G molecule on a BN nanosheet without deformation (initial state), along with the corresponding height mapping; (d-f) side and top views of a R6G molecule on the same BN nanosheet after conformational change (equilibrium state), along with the corresponding height mapping; (g-i) the distribution of strain e11, e12, and e22 in the deformed BN nanosheet.

Next, we conducted experiments to confirm the theoretical findings. The orientation of R6G adsorbed on BN nanosheets or graphene has never been experimentally analyzed, so angular dependence of near-edge X-ray adsorption fine structure (NEXAFS) spectroscopy was used to probe the orientation of R6G molecules adsorbed on a monolayer BN nanosheets. Before the adsorption of R6G, the BN nanosheet was cleaned by annealing in ultra-high vacuum (~$10^{-10}$ mba) at 425 ºC for 5 h. According to *in situ* X-ray photoelectron spectroscopy (XPS), the surface of the BN nanosheet was almost free of contamination after annealing (see Supporting Information, Figure S2). Figure 2a shows the C K-edge NEXAFS spectra of a monolayer of R6G molecules (9 Å thick) adsorbed on the clean BN nanosheet at different X-ray incident angles (*θ*). The π* resonances of the R6G molecules were highly angularly dependent, indicating that the orientation of the R6G on BN was not random. To determine the exact molecular orientation of the R6G on the BN, six peaks were used to fit the fine structures in the π* region, representing different C 1s-π* transitions in the molecules: 284.7 eV for the C 1s-$\pi^*_{C=C}$ transition of the aromatic carbons connected to the methyl groups,[15] 285.3 and 285.8 eV for the C 1s-$\pi^*_{C=C}$ transition of various aromatic carbons,[16] 286.4 eV for the C 1s-$\pi^*_{C=C}$ transition of the aromatic carbon bond to iminium groups,[15, 17] 287.6 eV for the C 1s-$\pi^*_{C=C}$ transition of the two carbon atoms adjacent to the oxygen atom in the tetrahydropyran ring,[16b] and 288.25 eV for the C 1s-$\pi^*_{C=O}$ transition of the carboxyl group.[16b] The C atoms corresponding to the six transitions are labeled in Figure 2b (C1-C6).

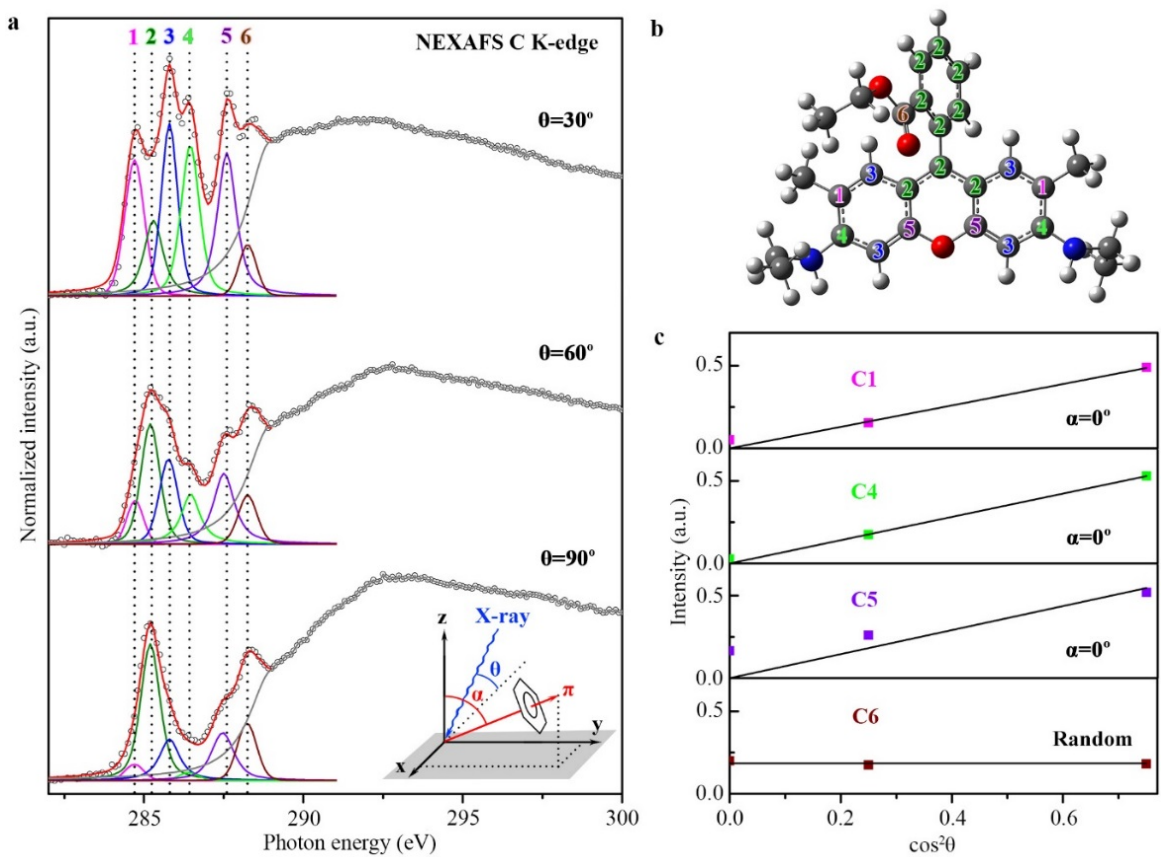

**Figure 2.** (a) Fittings to the π* resonance of the angularly-dependent NEXAFS spectra of the 9 Å R6G deposited on the BN nanosheet, with the geometry of the incidence shown in the inset;

(b) the C atoms corresponding to the six fitted π* transition sub-peaks in (a); (c) cosine-squared dependence of the intensities of the fitted C1, C4, C5 and C6 1s-π* transitions.

For a threefold substrate symmetry, the intensity of π* resonance (*I*) depends on $\theta$ and $\alpha$:[18]

$$I(\theta, \alpha) = A\left(cos^2\theta cos^2\alpha + \frac{1}{2}sin^2\theta sin^2\alpha\right) \quad (1)$$

where $\theta$ is X-ray incident angle; *A* is a constant; and $\alpha$ is the angle between the π vector perpendicular to an aromatic ring and the z-axis (inset in Figure 2a). For C1 and C4, the deduced $\alpha$ was close to 0°, and their intensities showed an excellent linear cosine-squared dependence (Figure 2c), implying the xanthene ring structure in R6G was parallel to the in-plane surface of BN, whose $\alpha \approx 0°$ (Figure 1).[19] The intensities of C5 slightly deviated from the cosine-squared fitting line at the different angles of incidence, possibly due to disturbance of the π orbitals by the oxygen atom in the ring system. In contrast, the intensity of C6 showed no polarization dependence, which is in accord with the rotational and hence randomly oriented C6 atom in the carboxyl structure. Therefore, R6G molecules were lying down on BN nanosheets, as shown in the calculation.

The conformational change in atomically thin BN after the coverage of R6G could not be visualized with AFM, so we propose to use Raman spectroscopy to experimentally detect them. Figure 3a (left) compares the G band ($E_{2g}$ mode) frequency of BN nanosheets suspended over pre-fabricated holes in a $SiO_2$/Si substrate before and after the immersion in $10^{-3}$ M R6G for 60 s, and the inset shows the corresponding optical microscopy image. The Raman spectrum of the as-prepared suspended BN nanosheet showed a G band centered at 1366.6 cm$^{-1}$, close to the value of bulk hBN crystals (~1366.4 cm$^{-1}$). In our previous study,[12c] we showed that atomically thin BN on $SiO_2$/Si had upshifted Raman frequency with decreased thickness, which was due to strain caused by the substrate. However, the suspended BN nanosheets were almost free of strain due to the absence of substrate's disturbance, and it suggests that the intrinsic Raman frequency of atomically thin BN is very close to that of bulk hBN. After R6G adsorption, the G band of the suspended nanosheet upshifted to 1368.3 cm$^{-1}$. The average frequency increase for 9 pieces of suspended BN nanosheets after the R6G adsorption was 1.6±0.6 cm$^{-1}$ (green arrow in Figure 3b). Interestingly, Raman shifts in the opposite direction were observed for BN nanosheets bound to SiO$_2$/Si after adsorption of R6G (middle of Figure 3a), with an average downshift of 1.3±0.7 cm$^{-1}$ from 8 nanosheets (blue arrow in Figure 3b). Note that these Raman shifts are more than one order of magnitude larger than the system error of the Raman measurements (see Supporting Information, Figure S4). In contrast, the G band frequency of bulk hBN crystals did not change before and after adsorption of R6G (right of Figure 3a). The Raman shifts were not caused by water, as in our control experiments, neither substrate-bound nor suspended BN nanosheets had Raman G band changes after immersed in water (without R6G). Several factors can cause Raman shift, but it is obvious that crystallinity and temperature factors are irrelevant in the current study. The observed Raman shifts cannot be attributed to doping by the adsorbed R6G molecules either, because unlike graphene, BN nanosheets are insulators and have negligible doping effect after molecule adsorption according to previous theoretical and experimental investigations.[9d, 14] As well, doping cannot explain the opposing Raman shifts observed on the suspended and substrate-bound BN nanosheets after adsorption of R6G.

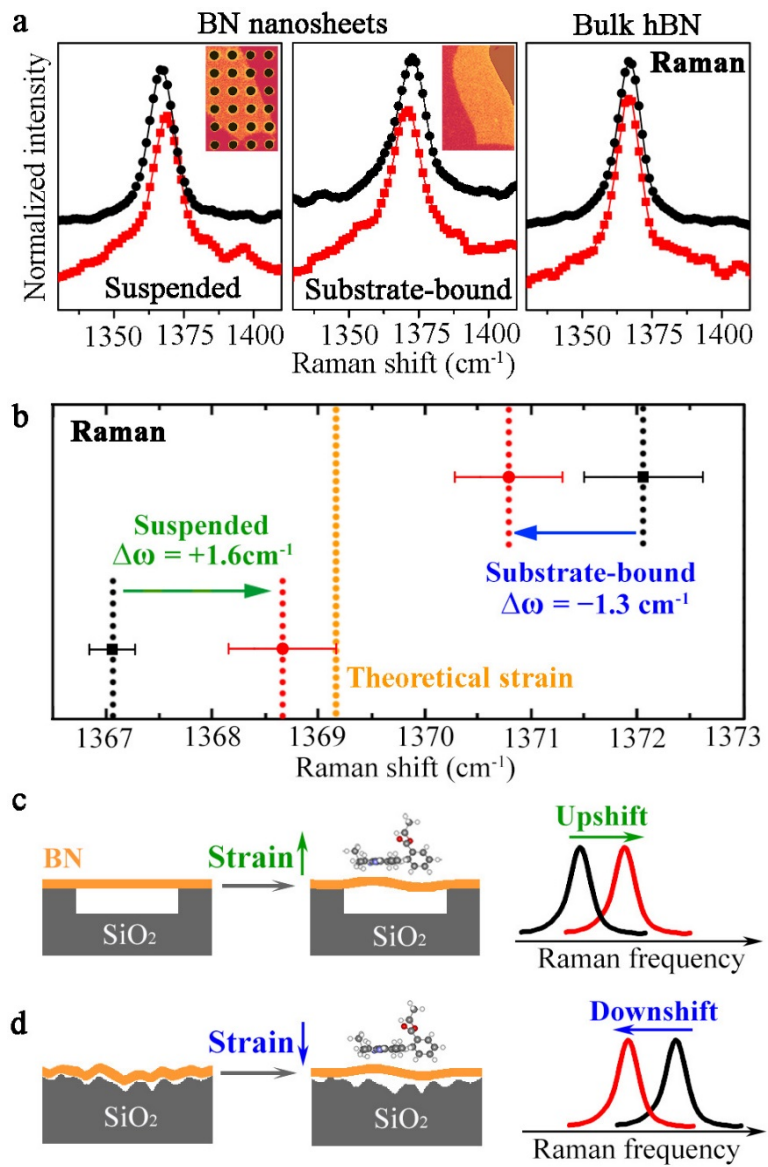

**Figure 3.** (a) Comparison of the Raman spectra of suspended and substrate-bound BN nanosheets as well as bulk hBN before (black) and after (red) adsorption of R6G molecules, with the optical images as insets; (b) summary of the average frequency shifts of the Raman G band before (black) and after (red) the adsorption of R6G on suspended ($N$=9) and substrate-bound ($N$=8) BN nanosheets, with the theoretically predicted strain labelled in gold; diagrams showing strain changes in (c) suspended and (d) substrate-bound BN nanosheets after adsorption of R6G, and the corresponding Raman shifts.

The observed opposite Raman shifts were caused by increased (decreased) compressive strain in the suspended (substrate-bound) BN nanosheets due to adsorption of R6G. A Raman shift derived from biaxial strain change in BN can be estimated by:[10c]

$$\Delta\varepsilon = -\Delta\omega_G/2\gamma_G\omega_G^0 \qquad (2)$$

where $\Delta\omega_G$ is the frequency shift of the Raman G band of BN caused by strain change $\Delta\varepsilon$; $\gamma_G$ is the Grüneisen parameter of hBN (0.89);[20] and $\omega_G^0$ is the G band frequency of unstrained BN. As shown in Figure 3c, the as-prepared suspended BN nanosheets had almost no strain, but the adsorption of R6G made the BN nanosheets deformed and compressively strained. Based on Eq.2 and the theoretically calculated strain (–0.16% as aforementioned), the adsorption of R6G should upshift the G band of BN nanosheets to 1369.2 cm$^{-1}$, as indicated by the dotted gold line in Figure 3b. On the contrary, the as-prepared substrate-bound BN nanosheets initially had a relatively large compressive strain due to the uneven SiO$_2$ substrate,[12c, 21] which was larger than the strain induced by the adsorption of R6G (Figure 3d). As the result, the R6G adsorption made the nanosheets less deformed, which reduced their strain and downshifted their Raman band. Such strain change is justified by the height distribution and hence roughness change in BN nanosheets bound to SiO$_2$ and after R6G adsorption (see Supporting Information, Figure S5). Therefore, R6G molecules may have been able to lift and then deform the BN nanosheets by overcoming the substrate adhesion. From an energy point of view, this is plausible. Although the interaction between BN nanosheets and SiO$_2$ substrate has not been measured before, the adhesion energy between graphene and SiO$_2$ was reported to be 0.24-0.31 J/m$^2$.[22] The adsorption energy between R6G and a BN nanosheet was calculated to be 1.55 eV, *i.e.*, 0.26 J/m$^2$, which is in the same range as the substrate-adhesion energy.

Conformational changes in atomically thin BN is not exclusive to the adsorption of R6G; 4-MBA molecules can also deform atomically thin BN nanosheets. According to theoretical calculations, a 4-MBA molecule spontaneously sank into a BN nanosheet from the deformation-free initial state to equillibrium state (Figure 4a-f). The conformational change caused an average strain of −0.087% in the BN nanosheet, and the strain distributions are shown in Figure 4g-i. Similarly, Raman could be used to detect such conformational changes. As with R6G, opposing Raman shifts were recorded for the suspended and substrate-bound BN after the adsorption of 4-MBA: the suspended BN nanosheets showed an average Raman upshift of 1.2±0.4 cm$^{-1}$ (green arrow in Figure 5), and the substrate-bound 2L BN showed an average Raman downshift of 2.9±0.6 cm$^{-1}$ (blue arrow in Figure 5). The upshifted Raman frequency of the suspended BN nanosheets after adsorption of 4-MBA was due to conformational changes which increased strain in the nanosheets; and substrate-bound nanosheets showed reduced strain and hence downshifted Raman frequency. In both cases, the Raman frequency shifted towards the theoretically calculated value labelled by the dotted line in gold (Figure 5). The adsorption energy for 4-MBA on a deformed BN nanosheet was 0.25 J/m$^2$, similar to that of R6G. That is, the surface adsorption of 4-MBA can also lift atomically thin BN nanosheets up from SiO$_2$/Si substrate. Molecule-induced conformational change is expected to be a phenomenon common to the interaction between molecules and other 2D nanomaterials which also have small rigidity or high flexibility.

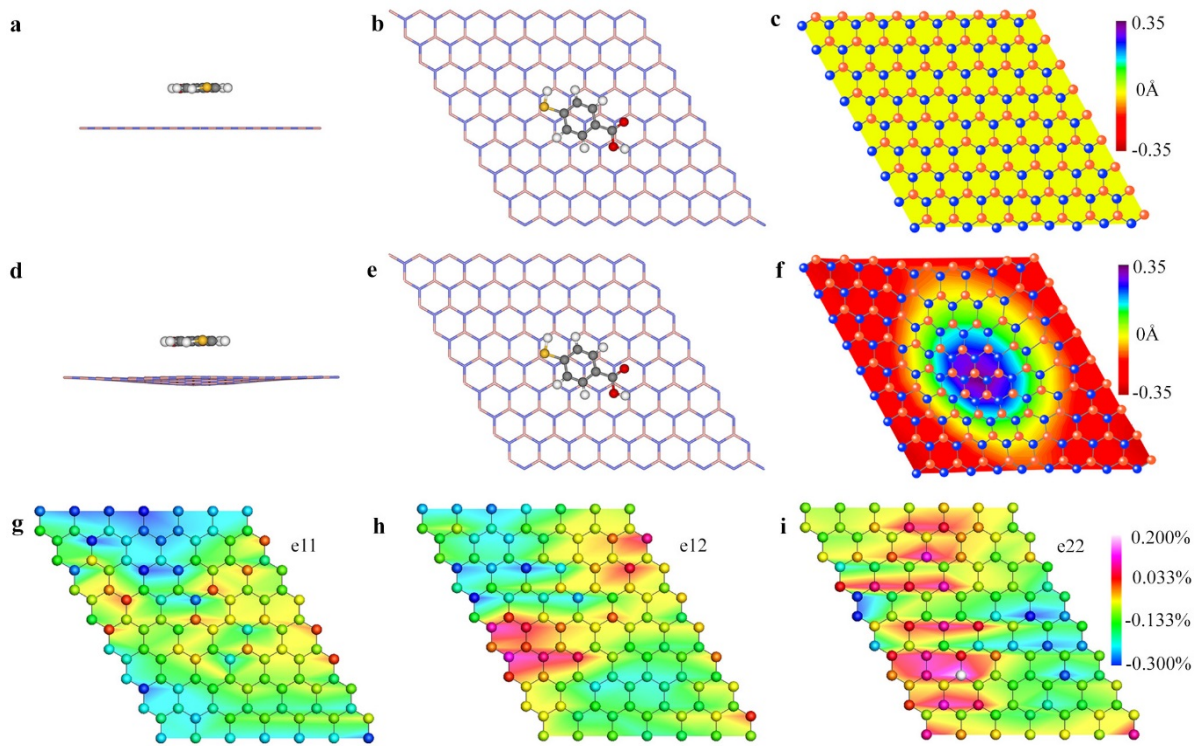

**Figure 4.** (a-c) Side and top views of a 4-MBA molecule on a BN nanosheet without deformation (initial state), along with the corresponding height mapping; (d-f) side and top views of a 4-MBA on the same BN nanosheet after conformational change (equilibrium state), along with the corresponding height mapping; (g-i) the distribution of strain e11, e12, and e22 in the deformed BN nanosheet.

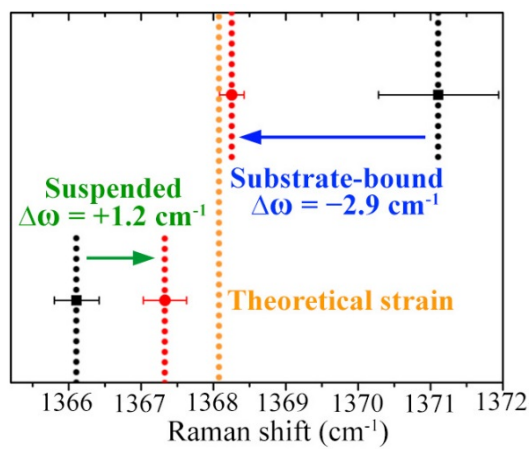

**Figure 5.** The frequency shifts of the Raman G band before (black) and after (red) the adsorption of 4-MBA molecules on suspended and substrate-bound BN nanosheets, with the theoretically predicted strain labelled in gold.

The special adsorption behavior in atomically thin nanosheets is valuable for many applications. For example, as aforementioned, conformational change can increase the adsorption energy and efficiency of nanosheets; in contrast, bulk crystals are not able to make such changes. In other words, nanosheets should have better surface adsorption than the corresponding bulk materials even with the same surface area. To compare the adsorption capability of atomically thin and bulk BN, we immersed BN nanosheets in an aqueous R6G solution ($10^{-3}$ M) for different lengths of time; the nanosheets were then washed with Milli-Q water to remove non-adsorbed molecules. After immersion for 40 s, 1L and 2L BN nanosheets were covered with densely packed islands of adsorbed R6G molecules with a thickness of ~0.8-1.0 nm (Figure 6a and b). This thickness corresponds to a monolayer of lying-down R6G, consistent with our DFT calculations. After 60 s, the R6G islands converged to form a complete monolayer, on top of which new islands started to form. After 300 s of immersion, relatively larger regions of 2-3 layers of R6G dominated the surface. Therefore, the physisorption of R6G on atomically thin BN seemed to follow the Stranski-Krastanov (layer-by-island) growth mode, as illustrated in Figure 6c. Less R6G adsorption occurred on 5L BN than on 1-3L BN (see Supporting Information, Figure S6). No complete layer of R6G was found on the surface of bulk hBN crystals even after 300 s of immersion; instead, only discontinuous islands of 2-layer-thick R6G were present (Figure 6d). Therefore, the adsorption on bulk hBN followed the Volmer-Weber (island) mode. These results suggest that the adsorption energy of R6G on atomically thin BN was greater than the cohesion energy among R6G molecules, whereas the adsorption energy of R6G on bulk hBN was less than the cohesion energy. That is, the adsorption energy of R6G on atomically thin BN and R6G is larger than that on bulk hBN, in full agreement with the expectation from conformational change. A similar trend in adsorption was obtained from the suspended BN nanosheets (see Supporting Information, Figure S7). The adsorption of 4-MBA on BN nanosheets was also superior to that on bulk hBN, as confirmed by AFM (Figure 7). Therefore, even with the same surface area, atomically thin BN is more effective than bulk hBN at adsorbing molecules due to conformational change.

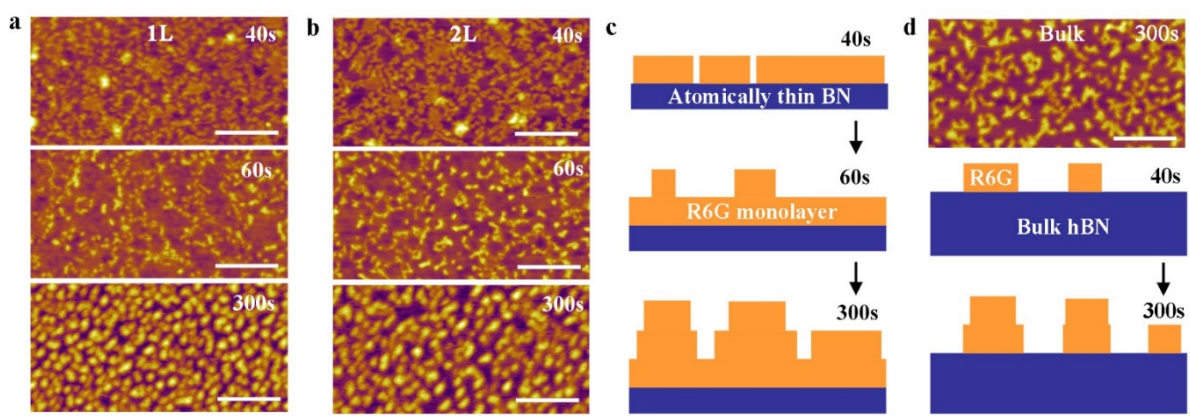

**Figure 6.** (a) 1L and (b) 2L BN after the immersion in aqueous solution of R6G ($10^{-3}$ M) for 40, 60 and 300 s (height range 5 nm); (c) schematic diagrams of the adsorption process of R6G on atomically thin BN; (d) AFM image of the R6G adsorbed on a bulk hBN crystal (~1 μm thick) under the same condition for 300 s (height range 5 nm) and the corresponding schematic diagrams. Scale bars in (a, b and d) 250 nm.

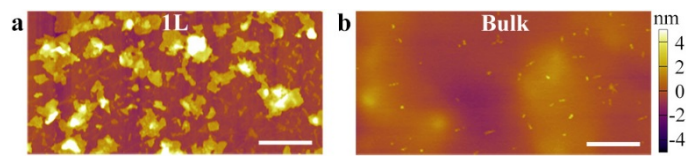

**Figure 7.** AFM images comparing the adsorption capability between a 1L and bulk BN on substrate (scale bars 500 nm).

To further demonstrate the applicaiton of conformational change, BN nanosheets and commercial (bulk) hBN particles of the same surface area were used for water cleaning. The nanosheets had a surface area 18 times larger than the bulk hBN particles (46.2 *v.s.* 2.6 m$^3$/g) as measured by BET method, so the cleaning effect of 3 mg of BN nanosheets towards R6G solution (10$^{-6}$ M) was compared to that of 54 mg of the hBN particles. After immersed in the solution for 5 min, the BN materials were removed by centrifugation, and the amounts of R6G left in the solution were compared. As shown in Figure 8a, BN nanosheets (Cycle 0) were more effective in removing R6G dyes from the solution than the bulk hBN particles, reflected by different color changes. According to the results from ultraviolet-visible (UV-Vis) spectroscopy in Figure 8b, the concentrations of R6G residues in the solution cleaned by BN particles and nanosheets were 3.7×10$^{-7}$ and 8.9×10$^{-8}$ M, respectively. That is, the bulk hBN particles removed ~60% of R6G from the solution; while BN nanosheets with the same surface area adsorbed ~90% of the dye due to the larger adsorption energy thanks to conformational change. These results are consistent with the AFM studies in Figure 6. In addition, the BN nanosheets can be reused without noticeable loss of efficiency. According to our previous study, atomically thin BN nanosheets are much more thermally stable than graphene and can survive ~800 °C in air.[12c] Therefore, the adsorbed organic molecules can be eliminated by heat treatment, and the regenerated BN nanosheets can be reused. It can be seen in Figure 8 (Cycle 1 and 2) that the water cleaning performance of the BN nanosheets after 2 cycles of reusability test by heating at 400 °C in air for 10 min was on a par with the starting material (Cycle 0). This experiment highlights the importance of thickness of a material besides its surface area and chemistry in water cleaning.

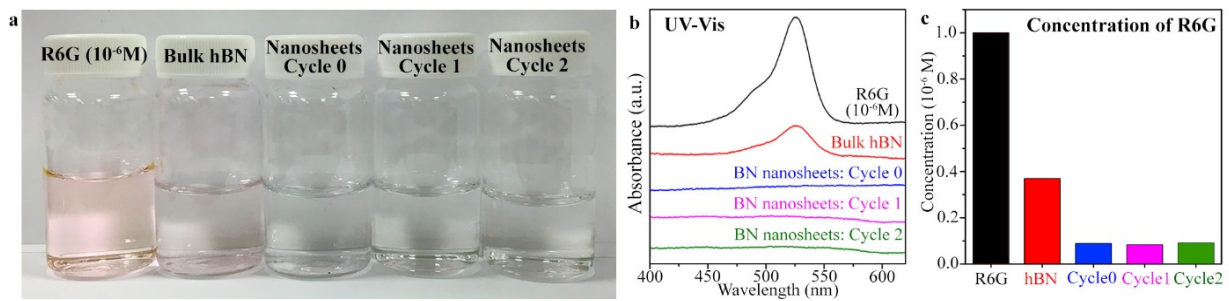

**Figure 8.** Photo of initial R6G water solution ($10^{-6}$ M) (far left) and after cleaning by bulk hBN particles (left), BN nanosheets (middle), and reused BN nanosheets for up to 2 cycles (right and far right); (b) UV-Vis spectra of the 5 solutions; (c) the estimated concentration of R6G residues in the solution after the cleaning by bulk hBN particles and BN nanosheets after different reusability cycles.

The conformational change induced stronger surface adsorption in BN nanosheets is also highly desirable for sensing. For this application, the BN nanosheets were added to R6G solution ($10^{-7}$ M), and the mixutre was dropped on pre-fabricated substrates for surface enhanced Raman spectroscopy (SERS). The SERS substrates were homogeneiously covered by plasmonically active silver nanoparticles. BN nanosheets could attract a large number of R6G molecules on their two surfaces and turned to red. When deposited on the SERS substrates, these concentrated or enriched molecules on BN nanosheets gave rise to strong Raman signals (black in Figure 9). In contrast, when BN nanosheets were not added, pure R6G solution showed very weak signals (red in Figure 9) because R6G molecules in this case were much diluted. No meaningful signals were detected from the bulk hBN particles after immersion under the same conditions likely due to their large thickness that blocked Raman scattering.

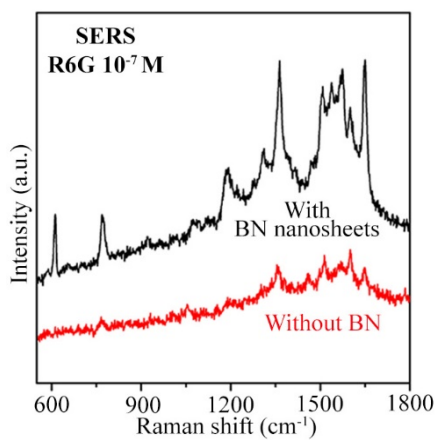

**Figure 9.** Detection of R6G ($10^{-7}$ M) by adsorption on BN nanosheets (black) and by direct deposition on SERS substrates (red).

## 3. Conclusion

DFT calculations showed conformational changes in atomically thin BN nanosheets after physisorption of molecules, such as R6G and 4-MBA, resulting in greater adsorption energy of the nanosheets. We experimentally detected such conformational changes in both suspended and substrate-bound BN nanosheets using Raman spectroscopy and NEXAFS. Furthermore, we found that atomically thin and bulk BN followed different adsorption modes, suggesting stronger adsorption capability of nanosheets. These results were fully consistent with the presence of conformational change in atomically thin nanosheets predicted by the theoretical calculations. We also demonstrated that the unique adsorption behavior of BN nanosheets led to many novel applications, including more efficient water cleaning and highly-sensitive molecule detection.

## 4. Experimental Section

DFT calculations were carried out using the Vienna Ab-initio Simulation Package (VASP). The exchange-correlation interaction was described by generalized gradient approximation (GGA) with the Perdew-Burke-Ernzerhof (PBE) functional. The van der Waals interactions were described by a DFT-D3 method with Becke-Jonson damping in all calculations. The lattice constants a and b were free to change during the relaxation while the lattice constant c was kept to 25 Å. The energy cut-off was set to 500 eV. Only the gamma point was used. The strain was computed based on the definitions of Green-Lagrangian strain.

The BN nanosheets on 90 nm $SiO_2$/Si substrate were mechanically exfoliated from single-crystal hBN by Scotch tape technique.[3c, 12c, 23] In detail, the hBN crystals were repeatedly exfoliated on a piece of tape, and then the tape with hBN was stuck onto the clean $SiO_2$/Si wafers. The manual removal of the tape left BN nanosheets on the substrates. Atomically thin BN was located using an Olympus optical microscope (BX51) and the thickness was measured by a Cypher AFM in both tapping and contact modes. The average thicknesses for the 1L, 2L, and 3L BN nanosheets were 0.43, 0.89 and 1.26 nm, respectively (Supporting Information, Figure S8). Suspended BN nanosheets were exfoliated on pre-fabricated 90 nm $SiO_2$/Si substrate with 1.2 μm holes (depth: 2 μm) following the same exfoliation procedure. Raman spectra were taken with a Renishaw inVia Raman microscope equipped with a 514.5 nm laser. A 100× objective lens with a numerical aperture of 0.90 was used. The laser power was ~2.5 mW. All Raman spectra were calibrated with the Raman band of Si at 520.5 $cm^{-1}$. Compared with bulk hBN crystals whose Raman G band was at 1366.4 $cm^{-1}$, atomically thin BN on $SiO_2$/Si showed upshifted G band frequencies due to the disturbance from the substrate and hence compressive strain in the nanosheets.[3c, 12c]

The BN nanosheets for water cleaning and sensing were synthesized by chemical blowing method.[24] They were few-layer thick (see Supporting Information, Figure S9). The as-synthesized BN nanosheets (3mg) and commercial bulk hBN particles (54mg) were immersed in R6G aqueous solution (50 mL@$10^{-6}$ M) for 5 min. After centrifugation at 14,000 rpm for 10 min to remove BN, the solution was collected for UV-Vis spectroscopy. The UV-Vis spectra was collected from a Varian Cary 300 spectrometer under double beam mode. The scan rate was 600 nm/min, and the scan step was 1.000 nm. For reusability test, the BN nanosheets with R6G were heated at 400 °C in air for 10 min to burn out the adsorbed molecules. For sensing, the BN nanosheets were firstly dispersed in water by sonication, and then 2-3 mL of the solution was added to $10^{-7}$ M R6G solution. The red sediment of BN nanosheets (with R6G adsorption) was deposited on pre-fabricated SERS substrate. The SERS substrates were produced by sputtering of a 10 nm Ag film on $SiO_2$/Si wafer and then heating at 500 °C in Ar to convert the film to plasmonic nanoparticles.[3c, 3g, 3j] For the SERS of pure R6G solution without BN nanosheets, same volume of the R6G solution was dropped on the SERS substrate.

**Supporting Information**

Supporting Information is available from the Wiley Online Library or from the author.

**Acknowledgements**

L.H.Li thanks Abu Sadek for the fabrication of the percolated Si wafer. Financial support from Australian Research Council via Discovery Early Career Researcher Award (DE160100796) and Discovery Project, Deakin University via ADPRF2014 and CRGS2015, and NSFC No.51420105002 is acknowledged. This research was partly undertaken on the soft x-ray beamline at the Australian Synchrotron, Victoria, Australia and at the Melbourne Centre for Nanofabrication (MCN) in the Victorian Node of the Australian National Fabrication Facility (ANFF).

[1] a) S. L. Burkett, M. J. Read, *Langmuir* **2001**, *17*, 5059; b) P. Roach, D. Farrar, C. C. Perry, *J. Am. Chem. Soc.* **2005**, *127*, 8168.

[2] S. Weigelt, C. Busse, L. Petersen, E. Rauls, B. Hammer, K. V. Gothelf, F. Besenbacher, T. R. Linderoth, *Nat. Mater.* **2006**, *5*, 112.

[3] a) P. Dai, Y. Xue, X. Wang, Q. Weng, C. Zhang, X. Jiang, D. Tang, X. Wang, N. Kawamoto, Y. Ide, *Nanoscale* **2015**, *7*, 18992; b) Y. Lin, C. E. Bunker, K. A. S. Fernando, J. W. Connell, *ACS Appl. Mater. Interfaces.* **2012**, *4*, 1110; c) Q. Cai, L. H. Li, Y. Yu, Y. Liu, S. Huang, Y. Chen, K. Watanabe, T. Taniguchi, *Phys. Chem. Chem. Phys.* **2015**, *17*, 7761; d) J. Li, X. Xiao, X. W. Xu, J. Lin, Y. Huang, Y. M. Xue, P. Jin, J. Zou, C. C. Tang, *Sci. Rep.* **2013**, *3*, 3208; e) Y. L. Yu, H. Chen, Y. Liu, V. Craig, L. H. Li, Y. Chen, *Adv. Mater. Interfaces* **2014**, *1*, 1300002; f) Y. L. Yu, H. Chen, Y. Liu, V. S. J. Craig, C. M. Wang, L. H. Li, Y. Chen, *Adv. Mater. Interfaces* **2015**, *2*, 1400267; g) Q. Cai, S. Mateti, W. Yang, R. Jones, K. Watanabe, T.

Taniguchi, S. Huang, Y. Chen, L. H. Li, *Angew. Chem. Int. Ed.* **2016**, *55*, 8405; h) L. H. Li, Y. Chen, *Adv. Funct. Mater.* **2016**, *26*, 2594; i) W. Sun, Y. Meng, Q. Fu, F. Wang, G. Wang, W. Gao, X. Huang, F. Lu, *ACS Appl. Mater. Interfaces.* **2016**, *8*, 9881; j) Q. Cai, S. Mateti, K. Watanabe, T. Taniguchi, S.-M. Huang, Y. Chen, L. H. Li, *ACS Appl. Mater. Interfaces* **2016**; k) C. Y. Zhi, Y. Bando, C. C. Tang, H. Kuwahara, D. Golberg, *Adv. Mater.* **2009**, *21*, 2889; l) Z. Wang, Z. Tang, Q. Xue, Y. Huang, Y. Huang, M. Zhu, Z. Pei, H. Li, H. Jiang, C. Fu, C. Zhi, *Chem. Rec.* **2016**, *16*, 1204.

[4] a) J. C. Meyer, A. K. Geim, M. I. Katsnelson, K. S. Novoselov, T. J. Booth, S. Roth, *Nature* **2007**, *446*, 60; b) Z. P. Xu, M. J. Buehler, *ACS Nano* **2010**, *4*, 3869; c) D. B. Zhang, E. Akatyeva, T. Dumitrica, *Phys. Rev. Lett.* **2011**, *106*, 255503.

[5] A. R. Koltonow, J. Huang, *Science* **2016**, *351*, 1395.

[6] S. A. Svatek, O. R. Scott, J. P. H. Rivett, K. Wright, M. Baldoni, E. Bichoutskaia, T. Taniguchi, K. Watanabe, A. J. Marsden, N. R. Wilson, P. H. Beton, *Nano Lett.* **2015**, *15*, 159.

[7] A. C. Ferrari, J. C. Meyer, V. Scardaci, C. Casiraghi, M. Lazzeri, F. Mauri, S. Piscanec, D. Jiang, K. S. Novoselov, S. Roth, A. K. Geim, *Phys. Rev. Lett.* **2006**, *97*, 187401.

[8] a) L. G. Cancado, A. Jorio, E. H. M. Ferreira, F. Stavale, C. A. Achete, R. B. Capaz, M. V. O. Moutinho, A. Lombardo, T. S. Kulmala, A. C. Ferrari, *Nano Lett.* **2011**, *11*, 3190; b) T. Xing, L. H. Li, L. T. Hou, X. P. Hu, S. X. Zhou, R. Peter, M. Petravic, Y. Chen, *Carbon* **2013**, *57*, 515; c) A. C. Ferrari, *Solid State Commun.* **2007**, *143*, 47.

[9] a) M. Kalbac, A. Reina-Cecco, H. Farhat, J. Kong, L. Kavan, M. S. Dresselhaus, *ACS Nano* **2010**, *4*, 6055; b) A. Das, S. Pisana, B. Chakraborty, S. Piscanec, S. K. Saha, U. V. Waghmare, K. S. Novoselov, H. R. Krishnamurthy, A. K. Geim, A. C. Ferrari, A. K. Sood, *Nat. Nanotechnol.* **2008**, *3*, 210; c) J. E. Lee, G. Ahn, J. Shim, Y. S. Lee, S. Ryu, *Nat. Commun.* **2012**, *3*, 1024; d) Z. Y. Chen, P. Darancet, L. Wang, A. C. Crowther, Y. D. Gao, C. R. Dean, T. Taniguchi, K. Watanabe, J. Hone, C. A. Marianetti, L. E. Brus, *ACS Nano* **2014**, *8*, 2943; e) X. C. Dong, D. L. Fu, W. J. Fang, Y. M. Shi, P. Chen, L. J. Li, *Small* **2009**, *5*, 1422; f) H.

Medina, Y. C. Lin, D. Obergfell, P. W. Chiu, *Adv. Funct. Mater.* **2011**, *21*, 2687; g) M. Bruna, A. K. Ott, M. Ijas, D. Yoon, U. Sassi, A. C. Ferrari, *ACS Nano* **2014**, *8*, 7432.

[10]   a) M. Y. Huang, H. G. Yan, C. Y. Chen, D. H. Song, T. F. Heinz, J. Hone, *Proc. Natl. Acad. Sci. U. S. A.* **2009**, *106*, 7304; b) C. Neumann, S. Reichardt, P. Venezuela, M. Drogeler, L. Banszerus, M. Schmitz, K. Watanabe, T. Taniguchi, F. Mauri, B. Beschoten, S. V. Rotkin, C. Stampfer, *Nat. Commun.* **2015**, *6*, 8429; c) T. M. G. Mohiuddin, A. Lombardo, R. R. Nair, A. Bonetti, G. Savini, R. Jalil, N. Bonini, D. M. Basko, C. Galiotis, N. Marzari, K. S. Novoselov, A. K. Geim, A. C. Ferrari, *Phys. Rev. B* **2009**, *79*, 205433; d) Z. H. Ni, T. Yu, Y. H. Lu, Y. Y. Wang, Y. P. Feng, Z. X. Shen, *ACS Nano* **2008**, *2*, 2301; e) W. Pan, J. L. Xiao, J. W. Zhu, C. X. Yu, G. Zhang, Z. H. Ni, K. Watanabe, T. Taniguchi, Y. Shi, X. R. Wang, *Sci. Rep.* **2012**, *2*, 893; f) J. Zabel, R. R. Nair, A. Ott, T. Georgiou, A. K. Geim, K. S. Novoselov, C. Casiraghi, *Nano Lett.* **2012**, *12*, 617; g) D. Yoon, Y.-W. Son, H. Cheong, *Phys. Rev. Lett.* **2011**, *106*, 155502; h) X. C. Dong, Y. M. Shi, Y. Zhao, D. M. Chen, J. Ye, Y. G. Yao, F. Gao, Z. H. Ni, T. Yu, Z. X. Shen, Y. X. Huang, P. Chen, L. J. Li, *Phys. Rev. Lett.* **2009**, *102*, 135501.

[11]   a) I. Calizo, A. A. Balandin, W. Bao, F. Miao, C. N. Lau, *Nano Lett.* **2007**, *7*, 2645; b) A. A. Balandin, S. Ghosh, W. Z. Bao, I. Calizo, D. Teweldebrhan, F. Miao, C. N. Lau, *Nano Lett.* **2008**, *8*, 902; c) W. W. Cai, A. L. Moore, Y. W. Zhu, X. S. Li, S. S. Chen, L. Shi, R. S. Ruoff, *Nano Lett.* **2010**, *10*, 1645.

[12]   a) L. Song, L. J. Ci, H. Lu, P. B. Sorokin, C. H. Jin, J. Ni, A. G. Kvashnin, D. G. Kvashnin, J. Lou, B. I. Yakobson, P. M. Ajayan, *Nano Lett.* **2010**, *10*, 3209; b) R. V. Gorbachev, I. Riaz, R. R. Nair, R. Jalil, L. Britnell, B. D. Belle, E. W. Hill, K. S. Novoselov, K. Watanabe, T. Taniguchi, A. K. Geim, P. Blake, *Small* **2011**, *7*, 465; c) L. H. Li, J. Cervenka, K. Watanabe, T. Taniguchi, Y. Chen, *ACS Nano* **2014**, *8*, 1457; d) H. Q. Zhou, J. X. Zhu, Z. Liu, Z. Yan, X. J. Fan, J. Lin, G. Wang, Q. Y. Yan, T. Yu, P. M. Ajayan, J. M. Tour, *Nano Res.* **2014**, *7*, 1232.

[13]   a) E. Koo, S. Y. Ju, *Carbon* **2015**, *86*, 318; b) N. Peimyoo, T. Yu, J. Z. Shang, C. X. Cong, H. P. Yang, *Carbon* **2012**, *50*, 201; c) X. C. Dong, Y. M. Shi, P. Chen, Q. D. Ling, W.

Huang, *Jap. J. Appl. Phys.* **2010**, *49*, 1S; d) K. Gopalakrishnan, K. Moses, P. Dubey, C. N. R. Rao, *J. Mol. Struct.* **2012**, *1023*, 2; e) X. Q. Meng, S. Tongay, J. Kang, Z. H. Chen, F. M. Wu, S. S. Li, J. B. Xia, J. B. Li, J. Q. Wu, *Carbon* **2013**, *57*, 507; f) I. Jo, Y. Kim, J. Moon, S. Park, J. S. Moon, W. B. Park, J. S. Lee, B. H. Hong, *Phys. Chem. Chem. Phys.* **2015**, *17*, 29492; g) P. Nguyen, J. W. Li, T. S. Sreeprasad, K. Jasuja, N. Mohanty, M. Ikenberry, K. Hohn, V. B. Shenoy, V. Berry, *Small* **2013**, *9*, 3823; h) B. Das, R. Voggu, C. S. Rout, C. N. R. Rao, *Chem. Commun.* **2008**, 5155.

[14]   X. Ling, W. J. Fang, Y. H. Lee, P. T. Araujo, X. Zhang, J. F. Rodriguez-Nieva, Y. X. Lin, J. Zhang, J. Kong, M. S. Dresselhaus, *Nano Lett.* **2014**, *14*, 3033.

[15]   O. Plekan, V. Feyer, R. Richter, M. Coreno, M. De Simone, K. Prince, A. Trofimov, E. Gromov, I. Zaytseva, J. Schirmer, *Chem. Phys.* **2008**, *347*, 360.

[16]   a) E. Hollauer, E. d. S. Prucole, M. L. M. Rocco, A. D. P. Netto, A. Schöll, R. Fink, *J. Braz. Chem. Soc.* **2005**, *16*, 31; b) D. Solomon, J. Lehmann, J. Kinyangi, B. Liang, K. Heymann, L. Dathe, K. Hanley, S. Wirick, C. Jacobsen, *Soil Sci. Soc. Am. J.* **2009**, *73*, 1817.

[17]   C. Kolczewski, R. Püttner, O. Plashkevych, H. Ågren, V. Staemmler, M. Martins, G. Snell, A. Schlachter, M. Sant'Anna, G. Kaindl, *J. Chem. Phys.* **2001**, *115*, 6426.

[18]   J. Stöhr, *NEXAFS spectroscopy*, Springer-Verlag, Berlin; New York **1992**.

[19]   L. H. Li, M. Petravic, B. C. C. Cowie, T. Xing, R. Peter, Y. Chen, C. Si, W. H. Duan, *Appl. Phys. Lett.* **2012**, *101*, 191604.

[20]   S. K. Singh, M. Neek-Amal, S. Costamagna, F. M. Peeters, *Phys. Rev. B* **2013**, *87*, 184106.

[21]   a) M. Ishigami, J. H. Chen, W. G. Cullen, M. S. Fuhrer, E. D. Williams, *Nano Lett.* **2007**, *7*, 1643; b) W. G. Cullen, M. Yamamoto, K. M. Burson, J. H. Chen, C. Jang, L. Li, M. S. Fuhrer, E. D. Williams, *Phys. Rev. Lett.* **2010**, *105*, 215504; c) W. Pan, J. Xiao, J. Zhu, C. Yu, G. Zhang, Z. Ni, K. Watanabe, T. Taniguchi, Y. Shi, X. Wang, *Sci. Rep.* **2012**, *2*, 893.

[22]  a) Z. Zong, C. L. Chen, M. R. Dokmeci, K. T. Wan, *J. Appl. Phys.* **2010**, *107*, 026104; b) S. P. Koenig, N. G. Boddeti, M. L. Dunn, J. S. Bunch, *Nat. Nanotechnol.* **2011**, *6*, 543; c) N. G. Boddeti, S. P. Koenig, R. Long, J. L. Xiao, J. S. Bunch, M. L. Dunn, *J. Appl. Mech.* **2013**, *80*, 040909.

[23]  L. H. Li, E. J. Santos, T. Xing, E. Cappelluti, R. Roldán, Y. Chen, K. Watanabe, T. Taniguchi, *Nano Lett.* **2015**, *15*, 218.

[24]  X. B. Wang, C. Y. Zhi, L. Li, H. B. Zeng, C. Li, M. Mitome, D. Golberg, Y. Bando, *Adv. Mater.* **2011**, *23*, 4072.